\documentclass[fleqn,12pt]{article} 
\usepackage{url}
 \pdfoutput=1
\usepackage{graphicx}
\usepackage{epsfig} 
\usepackage{wrapfig}
\usepackage{amsmath}
\begin{document}

\begin{center}
\Large \bf {Two-point theory for the differential self-interrogation Feynman-alpha method}
\end{center}
\vspace{0.5cm}
\begin{center}
{Johan Anderson\footnote{johan@nephy.chalmers.se}, Dina Chernikova$^{1}$, Imre P\'{a}zsit$^{1,3}$}, L\'en\'ard P\'al$^{2}$ and Sara A. Pozzi$^{3}$
\end{center}
\begin{center}
$^1$ Department of Nuclear Engineering, Chalmers University of Technology, SE-41296, G\"{o}teborg, Sweden \\
\end{center}
\begin{center}
$^{2}$ Centre for Energy Research, Hungarian Academy of Sciences, H-1525 Budapest 114, POB 49, Hungary
\end{center}
\begin{center}
$^{3}$ Department of Nuclear Engineering and Radiological Sciences, University of
Michigan, Ann Arbor Michigan, USA
\end{center}


\begin{abstract}
\noindent  A Feynman-alpha formula has been derived in a two region domain pertaining the stochastic differential self-interrogation (DDSI) method and the differential die-away method (DDAA). Monte Carlo simulations have been used to assess the applicability of the variance to mean through determination of the physical reaction intensities of the physical processes in the two domains. More specifically, the branching processes of the neutrons in the two regions are described by the Chapman - Kolmogorov equation, including all reaction intensities for the various processes, that is used to derive a variance to mean relation for the process. The applicability of the Feynman-alpha or variance to mean formulae are assessed in DDSI and DDAA of spent fuel configurations.
\end{abstract}

\section{Introduction}
Special nuclear materials have been assessed using the differential die-away analysis (DDAA) method \cite{kunz1982, croft2003, jordanphd, jordan2007, jordan2008} for several decades. In this deterministic method an initial pulse of fast neutrons is injected to the sample and the time dependence of the detection rate of fast neutrons is used to determine the presence of fissile materials such as $^{235}U$ and $^{239}Pu$. Its application is suitable when the fissile material is embedded in a moderating surroundings such that the source neutrons induce thermal fission after having slowed down.

There exists, however, one significant drawback of pulsed measurements, that is the necessity of a neutron generator. The stochastic generalization of the DDAA method was recently suggested as an alternative method which eliminates the need for such an external source, namely the so-called Differential Die-away Self-Interrogation(DDSI) technique \cite{menlove09}. The DDSI method utilizes the inherent spontaneous neutron emission of the sample. In the absence of a trigger signal, the temporal decay of the correlations as a function of the time delay between two detections of fast neutrons is used in the DDSI method. This corresponds to a Rossi-alpha measurement with two energy groups.

An empirical two-group Rossi-alpha formula, representing the functional form of the DDSI formula, was derived in Ref. \cite{menlove09}. This result was subsequently re-derived from first principles with the use of the backward-type Kolmogorov equations \cite{a1,a2}. Note that, the two-group theory of the Rossi- and Feynman alpha formulae is interesting also in other areas than nuclear safeguards, such as pulsed and stationary source driven experiments measuring the reactivity in fast cores of accelerator driven sub-critical systems. In such experiments \cite{soule2004, munoz2011, berglof2011} it was found that two exponentials appear, indicating that the temporal behavior of the fast and thermal neutrons is separated, especially in fast reflected cores. This amplified the need for the two group versions of the Feynman and Rossi-alpha formulae \cite{a1,a2}.

Energy dependent aspects of neutron counting have been studied previously; however, most investigations are limited to formulae valid for one infinite region. Note that there exists exceptions, in Ref. \cite{kuramoto2007} a two region theory representing the core and a reflector region is discussed. In certain situations one cannot disregard the fact that the neutron processes are taking place in a non-homogeneous medium, which consists of different regions with different characteristics. For instance in the DDAA and DDSI methods, the fissile material is surrounded with a moderator. Hence the fission takes place in the inner region, whereas the slowing down in the outer one, and there is a current of neutrons between the two regions in both directions.

If the domain is divided into several regions, two or more distinct exponentials will appear in the Feynman- and Rossi alpha formulae. In this contribution, a stochastic theory for a branching process in a one group neutron population applied to two domains is studied, based on the previous results of Refs \cite{a1,a2,anderson2012}. In particular, we consider a counterpart of the Differential Self-interrogation Variance to Mean (DSVM) formula, that is derived by using the master equation or Kolmogorov forward approach \cite{anderson2012}. The model includes a spontaneous fission source of fast neutrons, absorption in both domains, thermal fission, and detection of fast neutrons. Furthermore, we have included the intensity of neutron passage from one region to the other. However, the present variance to mean formula is energy independent and in contrast to the energy-independent one point theory of multiplicity, where the only appearing parameter is the first collision probability, we find two distinct exponentials appearing.

The usefulness of the theoretically predicted variance to mean, that is based on the observation of the two exponentials depends on the quantitative values of the various within- and inter-region neutron reaction intensities, which appear as coefficients in the equations. To assess the applicability and expected performance of the DSVM method in practical situations, the above mentioned reaction intensities were determined from numerical Monte Carlo (MCNPX) simulations. Significance of different values of the reaction intensities of thermal and fast neutrons in the performance of the method is discussed.

\section{A two-point variance to mean formula}
In Ref. \cite{anderson2012}, the authors discussed the extension of the variance to mean or Feynman-alpha one-group to two-group theory in a stochastic setting described by the Kolmogorov forward approach. Here, we will instead consider a one-group theory extended to two domains, also known as two-point theory. In order to keep the possibility of analytical solutions, we will choose a description which still disregards the handling of the spatial transport. The model being considered is that of two adjacent homogeneous half-spaces with space-independent reaction intensities in both regions. The exchange of the neutrons between the two regions will be described by the two passage intensities into the two different directions. The philosophy behind this model is the same as in the two- or many-point theories of coupled reactor cores, first introduced by Avery \cite{ave58} where a two-region system is described by point kinetics in each region, whereas the interaction between the two regions is described by coupling constants. Obviously, the assumption of a generic passage intensities which are the same for all neutrons in the respective regions is only an approximation of the real case, and would only correspond to the physical situation if both regions were infinite and took up the same space by intertwining each other. Nevertheless, the approximation in neglecting the space dependence of the transition intensity is not significantly coarser than using one single infinite homogeneous model for the evaluation of measurements performed in finite systems.

In the model, we have included a Poisson source of fast neutrons described by the source strength $S_1$ that releases $n$ particles with probability $p_q(n)$ at an emission event (i.e. spontaneous fission) in region I. Furthermore, we assume that the source is switched on at time $t=t_0 < 0$, although the dependence on $t_0$ will not be denoted. The detection rate of particles is included and is denoted by the intensity $\lambda_d$. We will start by deriving the explicit analytical formulae describing all the possible processes in the system. The Kolmogorov forward equation will give us a differential equation for the probability $P(N_1, N_2, Z_1,t)$ for having $N_1$ neutrons in region I, $N_2$ neutrons in region II at time $t$ in the system and having detected $Z_1$ neutrons in the interval $(0,t)$. In deriving this differential equation we have summed all mutually exclusive events during an infinitesimally small time interval $dt$ and we find for the probability $P(N_1, N_2, Z_1,t)$ the differential equation
\begin{eqnarray}\label{eq:1.1}
\frac{\partial P(N_1,N_2,Z_1,t)}{\partial t} & = & - (\lambda_1 N_1 + \lambda_2 N_2 + S_1)P(N_1, N_2, Z_1,t) \nonumber \\
& + & \lambda_{1a} (N_1 + 1) P(N_1 + 1, N_2, Z_1,t) \nonumber \\
& + & \lambda_{2a} (N_2 + 1) P(N_1, N_2 + 1, Z_1,t) \nonumber \\
& + & \lambda_{1f} (N_1 + 1) \sum_k^{N_1} f^1_k P(N_1 - k, N_2, Z_1,t) \nonumber \\
& + & \lambda_{2f} (N_2 + 1) \sum_k^{N_2} f^2_k P(N_1, N_2 - k, Z_1,t) \nonumber \\
& + & \lambda_{T1} (N_2 + 1) P(N_1 - 1, N_2 + 1, Z_1, t) \nonumber \\
& + & \lambda_{T2} (N_1 + 1) P(N_1 + 1, N_2 - 1, Z_1, t) \nonumber \\
& + & \lambda_d (N_1 + 1) P(N_1 + 1, N_2, Z_1 - 1,t) \nonumber \\
& + & S_1 \sum_n^{N_1} p_q(n) P(N_1 - n, N_2, Z_1,t).
\end{eqnarray}
Here, $f^j_k$ is the probability of having exactly $k$ neutrons produced in an induced fission event in region $j$ and $\lambda_1$ and $\lambda_2$ are the decay constants (total reaction intensities) for particles in region I and II, respectively. The intensities $\lambda_{1a}$, $\lambda_{2a}$ are the absorption (actually, capture) intensities of region I and II, while fission resulting from the thermal particles happens with the intensity of $\lambda_{1f}$ and $\lambda_{2f}$. The intensity of the detection of a fast neutron is denoted by $\lambda_{d}$. The total intensities are given by
\begin{eqnarray}\label{eq:1.2}
\lambda_1 = \lambda_{1a} + \lambda_{1f} + \lambda_{T1} + \lambda_d,
\end{eqnarray}
and
\begin{eqnarray}\label{eq:1.3}
\lambda_2 = \lambda_{2a} + \lambda_{2f} + \lambda_{T2}.
\end{eqnarray}
Where, $\lambda_{T1}$ describes the intensity of particles leaving for region II and $\lambda_{T2}$ is the intensity of particles transferring to region I. We derive the equations for the factorial moments by using the generating function of the form
\begin{eqnarray}\label{eq:1.4}
G(X,Y,Z,t) & = & \sum_{N_1} \sum_{N_2} \sum_{Z_1} X^{N_1} Y^{N_2} Z^{Z_1} P(N_1,N_2,Z_1,t),
\end{eqnarray}
and describe the time evolution of the process by a partial differential equation in the variables $(X,Y,Z)$ in terms of the generating function as,
\begin{eqnarray}\label{eq:1.5}
\frac{\partial G}{\partial t} & = & (\lambda_{1a} + q^1(X) \lambda_{1f} + \lambda_d Z - \lambda_1 X - \lambda_{T1} Y) \frac{\partial G}{\partial X} \nonumber \\
& + & (\lambda_{2a}  + \lambda_{2f} q^2(Y) - \lambda_2 Y - \lambda_{T2} X) \frac{\partial G}{\partial Y} \nonumber \\
& + & S_1 (r(X) - 1) G,
\end{eqnarray}
where
\begin{eqnarray}
q^j(X) & = & \sum_k f^j_k X^k, \label{eq:1.61} \;\;\;\;\;\;\;\;\;\;  \mbox{and} \\
r(X) & = & \sum_n p_q(n) X^n. \label{eq:1.62}
\end{eqnarray}
Here the superscript $j$ denotes the possibilities of having different material compositions in region I and II, respectively. In the numerical analysis we will assume that $q^2(Y)$ vanishes signifying that region I is surrounded by moderating material without fissile components. We have used the definition of the derivatives of the expressions (\ref{eq:1.61}) and (\ref{eq:1.62}) as $\nu^j_1 = d q^j/dX|_{X=1}$ and $r_1 = d r/dX|_{X=1}$, which stand for the expectations of the number of neutrons from induced and sponaneous fissions, respectively \cite{pazpal08}. We note that for $t_0 \rightarrow -\infty$, the expectations of neutrons in the two regions ($\langle N_1 \rangle$) and ($\langle N_2 \rangle$) will reach steady state due to the stationary source term with intensity $S_1$. The solutions to the system of differential equations are found by differentiation of equation (\ref{eq:1.5}) with respect to ($X,Y,Z$) and then letting ($X = Y = Z = 1$). These read as
\begin{eqnarray}
\langle N_1 \rangle & = & \bar{N}_1 = \frac{(- \lambda_2  + \nu_1^2 \lambda_{2f}) S_1 r_1 }{\lambda_{T1}
\lambda_{T2} - (-\lambda_1 + \nu^1_1 \lambda_{1f})(-\lambda_2 + \nu^2_1 \lambda_{2f})} \nonumber \\
& = & \frac{\lambda_2 S_1 r_1 ( \lambda_2 - \nu^2_1 \lambda_{2f})}{\omega_1 \omega_2}, \label{eq:1.10}\\
\langle N_2 \rangle & = & \bar{N}_2 = \frac{\lambda_{T1} S_1 r_1}{\omega_1 \omega_2}, \label{eq:1.11} \\
\langle Z_1 \rangle & = & \varepsilon \lambda_{1f} \bar{N}_1 t, \label{eq:1.12}
\end{eqnarray}
where $\varepsilon = \lambda_d/\lambda_{1f}$ and we have used the additional definitions $\omega_1$ and $\omega_2$ as
\begin{eqnarray}
\omega_1 & = & \frac{1}{2}(\lambda_2 + \lambda_2 - (\nu^1_1 \lambda_{1f} + \nu^2_1 \lambda_{2f})) \nonumber \\
& + &\frac{1}{2} \sqrt{((\lambda_1 - \lambda_2) - (\nu^1_1 \lambda_{f1} - \nu^2_1 \lambda_{2f}))^2 + 4 \lambda_{T1}  \lambda_{T2}}, \label{eq:1.13} \\
\omega_2 & = & \frac{1}{2}(\lambda_2 + \lambda_2 - (\nu^1_1 \lambda_{1f} + \nu^2_1 \lambda_{2f})) \nonumber \\
& - &\frac{1}{2} \sqrt{((\lambda_1 - \lambda_2) - (\nu^1_1 \lambda_{f1} - \nu^2_1 \lambda_{2f}))^2 + 4 \lambda_{T1}  \lambda_{T2}}, \label{eq:1.14}
\end{eqnarray}
It can be noted that the expectation of the detections increases linearly with time and the number of neutrons in the second domain is directly determined by the transfer from the first to the second region $\lambda_{T1}$. The linear increase in the expectation of the detections is due to the fact that it is determined by integration of the expectation of the neutron number with respect to time. In order to find the variance of the detector counts we need to determine the second moments by yet another differentiation with respect to ($X,Y,Z$) followed by letting ($X = Y = Z = 1$). We find the variance of the detector counts through the relation $\sigma_Z^2 = \langle Z_1 \rangle + \mu_{ZZ}$ where the modified variance $\mu_{ZZ}$ is defined as $\mu_{ZZ} = \langle Z(Z-1) \rangle - \langle Z\rangle^2 = \sigma_{ZZ}^2 - \langle Z\rangle$ while in general we have $\mu_{X Y} = \langle X Y\rangle - \langle X \rangle \langle Y \rangle$. The differentiation procedure results in a system of six ordinary differential equations for the second order modified moments. The differentiation procedure gives a system of six dynamical equations of the modified second moments as
\begin{eqnarray}
\frac{\partial}{\partial t} \mu_{X X} & = &  2 ( - \lambda_1 + \nu^1_1 \lambda_{1f}) \mu_{X X} - 2 \lambda_{T2} \mu_{X Y}  + \nu^1_2 \lambda_{1f} \bar{N}_1 + S_1 r_2, \label{eq:1.16} \\
\frac{\partial}{\partial t} \mu_{X Y} & = & (-\lambda_1 - \lambda_2 + \nu^1_1 \lambda_{1f} + \nu^2_1 \lambda_{2f}) \mu_{X Y} - \lambda_{T1} \mu_{X X} - \lambda_{T2} \mu_{Y Y}, \label{eq:1.17}\\
\frac{\partial}{\partial t} \mu_{Y Y} & = & 2 (- \lambda_2 + \nu_1^2 \lambda_{2f}) \mu_{Y Y} -  2 \lambda_{T2}
 \mu_{X Y},\label{eq:1.18} \\
\frac{\partial}{\partial t} \mu_{X Z} & = & (- \lambda_1 + \nu^1_1 \lambda_{1f})\mu_{X Z} - \lambda_{T2} \mu_{Y Z} + \lambda_d \mu_{X X},\label{eq:1.19} \\
\frac{\partial}{\partial t} \mu_{Y Z} & = & (-\lambda_2 + \nu^2_1 \lambda_{2f}) \mu_{Y Z} - \lambda_{T1} \mu_{X Z}  + \lambda_d  \mu_{X Y}, \label{eq:1.20} \\
\frac{\partial}{\partial t} \mu_{ZZ} & = & 2\epsilon \lambda_{1f} \mu_{X Z}, \label{eq:1.21}
\end{eqnarray}
The equation system and its solution is rather analogous to the case of the Feynman-alpha equations in two-group theory and with one neutron energy group but including delayed neutrons as given in Ref. \cite{anderson2012} and in Ref. \cite{pazsit1999}, respectively. Although an analytical solution for the general time-dependent system of equations (\ref{eq:1.16}) - (\ref{eq:1.21}) would be hard to find, we note that in the stationary state the system breaks down into two systems such that the solution of the first three equations is independent from the second, such that the moments  $\langle \mu_{X X} \rangle = \bar{\mu}_{XX} $, $\langle \mu_{X Y} \rangle = \bar{\mu}_{XY} $ and $\langle \mu_{Y Y} \rangle = \bar{\mu}_{YY} $ are constants. The equations describing detected particles need to be solved retaining the full time evolution by e.g. Laplace transforms. Moreover, it is found that the sought moment $\langle \mu_{Z Z} \rangle $ is determined by quadrature of moment $\langle \mu_{X Z} \rangle$. We find the constant 2nd moments as,
\begin{eqnarray}
\bar{\mu}_{XX} & = & \frac{((\lambda_2 - \nu_1^2 \lambda_{2f})^2 + \omega_1 \omega_2)
(\lambda_{T2} \bar{N}_2 + S_1 r_2)}{2(\lambda_1 + \lambda_2 - \nu_1^1 \lambda_{1f} - \nu_1^2 \lambda_{2f})\omega_1 \omega_2}, \label{eq:1.22}\\
\bar{\mu}_{XY} & = &  \frac{(\lambda_2 - \nu_1^2 \lambda_{2f}) \lambda_{T1} (\lambda_{T2}
\bar{N}_2 + S_1 r_2)}{2(\lambda_1 + \lambda_2 - \nu_1^1 \lambda_{1f} - \nu_1^2 \lambda_{2f})\omega_1 \omega_2}, \label{eq:1.23} \\
\bar{\mu}_{YY} & = & \frac{\lambda_{T1}^2 (\lambda_{T2}
\bar{N}_2 + S_1 r_2)}{2(\lambda_1 + \lambda_2 - \nu_1^1 \lambda_{1f} - \nu_1^2 \lambda_{2f})\omega_1 \omega_2}. \label{eq:1.24}
\end{eqnarray}
Here we have used the notation $r_2 = d^2 r/dX^2|_{X=1}$ for the second factorial moment \cite{pazpal08}. The objective now is to solve (\ref{eq:1.19}) and (\ref{eq:1.20}) by Laplace transform methods and we find the transformed identity as,
\begin{eqnarray}
\tilde{\mu}_{XZ} = \frac{\nu_1 \lambda_d \lambda_{T2} \bar{\mu}_{XY}}{s H(s)} + \frac{(s+\lambda_2 - \nu_1^2 \lambda_{2f})\lambda_d \bar{\mu}_{XX}}{s H(s)} \label{eq:1.25}
\end{eqnarray}
with
\begin{eqnarray}
H(s) = s^2 + (\omega_2 + \omega_1)s + \omega_1 \omega_2. \label{eq:1.26}
\end{eqnarray}
Note that we have assumed that the initial values of the moments $\mu_{XZ}$ and $\mu_{YZ}$ were equal to zero at $t=0$ (at the start of the measurement), hence the roots of $H(s)$ determine the temporal behavior of the Feynman-alpha formula. Moreover, the solution has many similarities to that found in Refs. \cite{pazsit1999, anderson2012}. We determine the variance to mean or Feynman-alpha formula by utilizing the relation for the variance $\sigma_{ZZ}^2 = \langle Z_1 \rangle + \mu_{ZZ}$ and after some algebra we find,
\begin{eqnarray}
\frac{\sigma_{ZZ}^2(T)}{\langle Z_1 \rangle} = 1 + Y_1 (1 - \frac{1-e^{- \omega_1 T}}{\omega_1 T}) + Y_2 (1 - \frac{1-e^{- \omega_2 T}}{\omega_2 T}). \label{eq:1.27}
\end{eqnarray}
Here, the complete expressions for $Y_1$ and $Y_2$ are quite
lengthy, and are therefore given in the Appendix. However, it turns out that the sum $Y_0 = Y_1 + Y_2$ takes a
rather simple form that also determines the value of the
Feynman-alpha for large gate times $T \rightarrow \infty$ as,
\begin{eqnarray}
Y_0 =  Y_1 + Y_2 = \nu_2 \frac{\lambda_d (\lambda_2 - \nu_1^2 \lambda_{2f}) \lambda_{T1} \lambda_{T2}}
{\omega_1^2 \omega_2^2}.\label{eq:1.28}
\end{eqnarray}
In the next section we will consider some quantitative examples of the variance to mean formula in Equation (\ref{eq:1.27}).

\section{Quantitative assessment of the variance to mean formula}
\begin{figure}[ht!]
\includegraphics[bb=0.0in 0.0in 10in 3in, width= 15cm, height = 5cm]{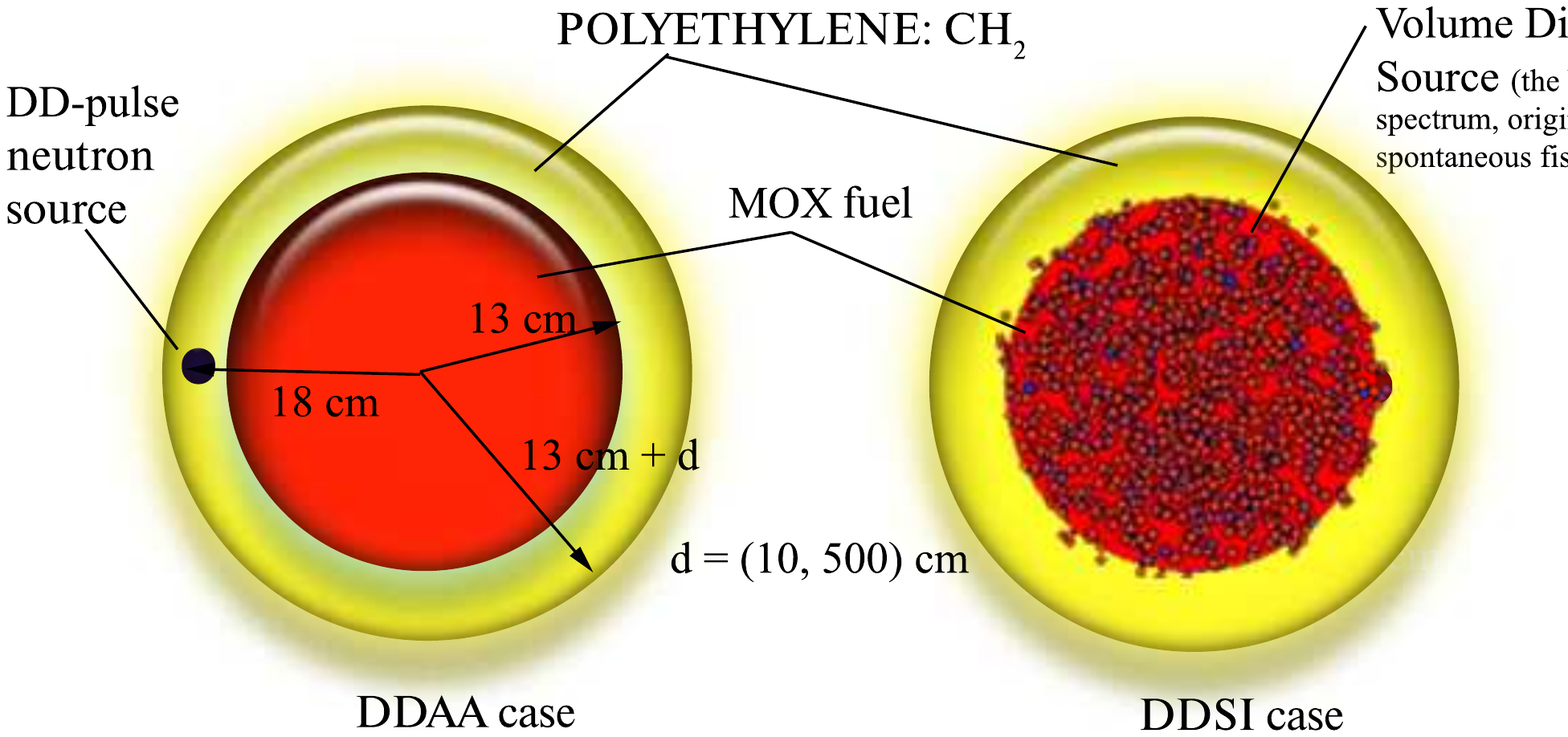}
\caption{Show the geometry used for the Monte Carlo (MCNPX) simulations
for the DDAA (left) and DDSI (right) cases in approximation of real and infinite moderator.}
\label{fig:fig1}
\end{figure}

We will now discuss the quantitative application of the previously found variance to mean (Eq. \ref{eq:1.27}) for a MOX fuel assembly, in DDAA and DDSI setups with varying amount of polyethylene moderator. In the simulations, the 58 fuel pins were recast into spherical geometry with the same amount of material, yielding a radius of roughly 13 cm. The spent fuel is of MOX type (U(94.14\%)O$_{2}$ + Pu(5.86\%)O$_{2}$) with approximately 4\% fissile content. We have considered two different thicknesses of the outer reflector shell with the sizes 10 and 500 cm where the larger moderator (500 cm) was assumed to be the equivalent of an infinite moderator. In the Monte-Carlo (MCNPX) simulations \cite{mcnpx} we consider two models, shown in Figure (\ref{fig:fig1}). The first model represents a DDAA case (left), where the external pulsed neutron source was placed at the distance of 5 cm far from fuel sample having the same energy of neutrons as those produced in DD-fusion reactions. Note, that there is no speciel significance to the selection of DD-fusion reactions, the DT-fusion would have yielded similar results with the appropriate changes to the moderator thickness. The second model was constructed to be similar to a DDSI case (right), where the source was assumed to be equally distributed in the fuel volume and the energies of the source neutrons were sampled from the Watt fission spectrum, originating from the spontaneous fission of $^{240}$Pu. The energy spectrum is determined by the function,
\begin{eqnarray} \label{eq:2.1}
p(E) \sim e^{-E/a} \sinh (b E)^{1/2}
\end{eqnarray}
where $a = 0.799$ MeV and $b = 4.903$ $MeV^{-1}$. Considering all possible cases we have hence carried out four MCNPX simulations in order to estimate the reaction intensities in two geometrical domains, fuel and moderator, with varying amount of moderation. In the simulations we have assumed a one energy group approximation. Reaction rates for each zone (and total) were determined by review of the neutron weight balance table (print table 130) of the MCNPX output which provides the fraction of source particles which enter each cell of the model, as it was done in papers of \cite{wmr2010} and \cite{broeder2000}. The simulation results are shown in Table 1 for the cases described above.
\begin{table}[ht]
\caption{Simulation Results} 
\centering 
\begin{tabular}{c c c c c c c} 
\hline\hline 
Intensity & DDAA (500cm) & DDAA (10cm) & DDSI (500cm) & DDSI (10cm) \\ [0.5ex] 

$\lambda_1$ $[1/s]$    & 0.7648   & 0.7441 & 1.6715 & 1.6665 &     \\
$\lambda_2$ $[1/s]$    & 1.5641   & 1.5502 & 2.0527 & 2.0453 &     \\
$\lambda_{T1}$ $[1/s]$ & 0.5641   & 0.5501 & 1.2422 & 1.2406 &     \\
$\lambda_{T2}$ $[1/s]$ & 0.4535   & 1.0185 & 0.8105 & 1.2588 &     \\
$\lambda_f$ $[1/s]$    & 0.1084   & 0.1049 & 0.2335 & 0.2317 &
\end{tabular}
\label{table:simulat} 
\end{table}

\begin{table}[ht]
\caption{Results} 
\centering 
\begin{tabular}{c c c c c c c} 
\hline\hline 
   & DDAA (500cm) & DDAA (10cm) & DDSI (500cm) & DDSI (10cm) \\ 
 $\nu_{eff}$          & 0.162    & 0.159  & 0.176 & 0.175    \\
 $\omega_1$ $[1/s]$   & 0.6318   & 0.5299 & 0.9578 & 0.8360  \\
 $\omega_2$ $[1/s]$   & 1.3837   & 1.4614 & 2.0916 & 2.1987
\end{tabular}
\label{table:simulat2} 
\end{table}
The simulation results are shown in Table \ref{table:simulat} for the cases described above, where the reaction intensities obtained from the simulations were normalized to one starting neutron having a relative error less than 4\%. In the interpretation of the simulation data, the sources in each region have to be properly accounted for remembering the specific setup of our analytical formula to achieve relevant comparisons. Withouth loss of generality in this study, we assume that there is no fission in the second region that contains only moderating material, by letting ($\lambda_{1f} = \lambda_f$, $\nu_1^1 = \nu_1$, $\lambda_{2f} = 0 = \nu_1^2$) and thus significantly simplifying the analytical formulas. In general we find a total balance of the intensities reflecting all the processes in the simulation (neglecting (n,xn) reactions that are several orders of magnitude smaller),
\begin{eqnarray} \label{eq:2.2}
1 + N_{fission} = N_{ltf} + N_{a 1} + N_{a 2} + N_{lost}.
\end{eqnarray}
Here, on the left hand side the source is represented by $1$ and $N_{fission}$, where $N_{fission}$ is the number of fissions whereas the losses are represented by particles lost in the fission process $N_{ltf}$ signifying the number of fission events (the actual number of particles lost to fission) and the absorbtions in the two regions $N_{a 1}$ and $N_{a 2}$, respectively. Furthermore, since the moderating material is finite, particles that may leave the system are accounted for by the term $N_{lost}$. Note that all $N$s are normalized to the number of source neutrons. Now we will consider the DDAA and DDSI situations separately as well as the two regions. In the DDSI situation we have a source in region I. For that region we have the balance,
\begin{eqnarray} \label{eq:2.3}
1 + N_{fission} = N_{ltf} + N_{a 1} + N_{T^{\prime}},
\end{eqnarray}
where $N_{T^{\prime}} = N_{I \rightarrow II} - N_{II \rightarrow I}$ is the total flux of neutrons from region I to II and in region II we find
\begin{eqnarray} \label{eq:2.4}
N_{I \rightarrow II} - N_{II \rightarrow I} = N_{a 2} + N_{lost}.
\end{eqnarray}
In the DDAA case we find slightly different balance equations of the form in region I,
\begin{eqnarray} \label{eq:2.5}
N_{fission} - N_{T^{\prime}} = N_{ltf} + N_{a 1},
\end{eqnarray}
and in region II
\begin{eqnarray} \label{eq:2.6}
1 + N_{T^{\prime}} = N_{a 2} + N_{lost}.
\end{eqnarray}
Here the source in region II is $1 + N_{T^{\prime}}$. In both the DDSI and DDAA cases, we identify $N_{T^{\prime}}$ with the net flux out of region I. Using the balance equations (\ref{eq:2.2}) - (\ref{eq:2.6}), we find the reaction intensities presented in Table \ref{table:simulat}.
In order to check the consistency of our numerical studies we computed the effective multiplication factor $\nu_{eff}$. In the simulated cases this value can be found by using the definition, the ratio of number of neutrons to the neutrons in the preceding generation,
\begin{eqnarray} \label{eq:2.7}
\nu_{eff} = \frac{\nu \lambda_f}{\lambda_1 + \lambda_2},
\end{eqnarray}
where the probability of fission is the fission intensity divided by the total intensities $\lambda_f/(\lambda_1 + \lambda_2)$. In combination with the multiplication factor $\nu_{eff}$ we compute the time constant $(\omega_1,\omega_2)$ describing the system, and the values are given in Table \ref{table:simulat2}.
\begin{figure}[ht!]
\begin{minipage}[b]{0.5\linewidth}
\centering
\includegraphics[bb=0.0in 3.0in 10in 8in, width=7cm, height = 4cm]{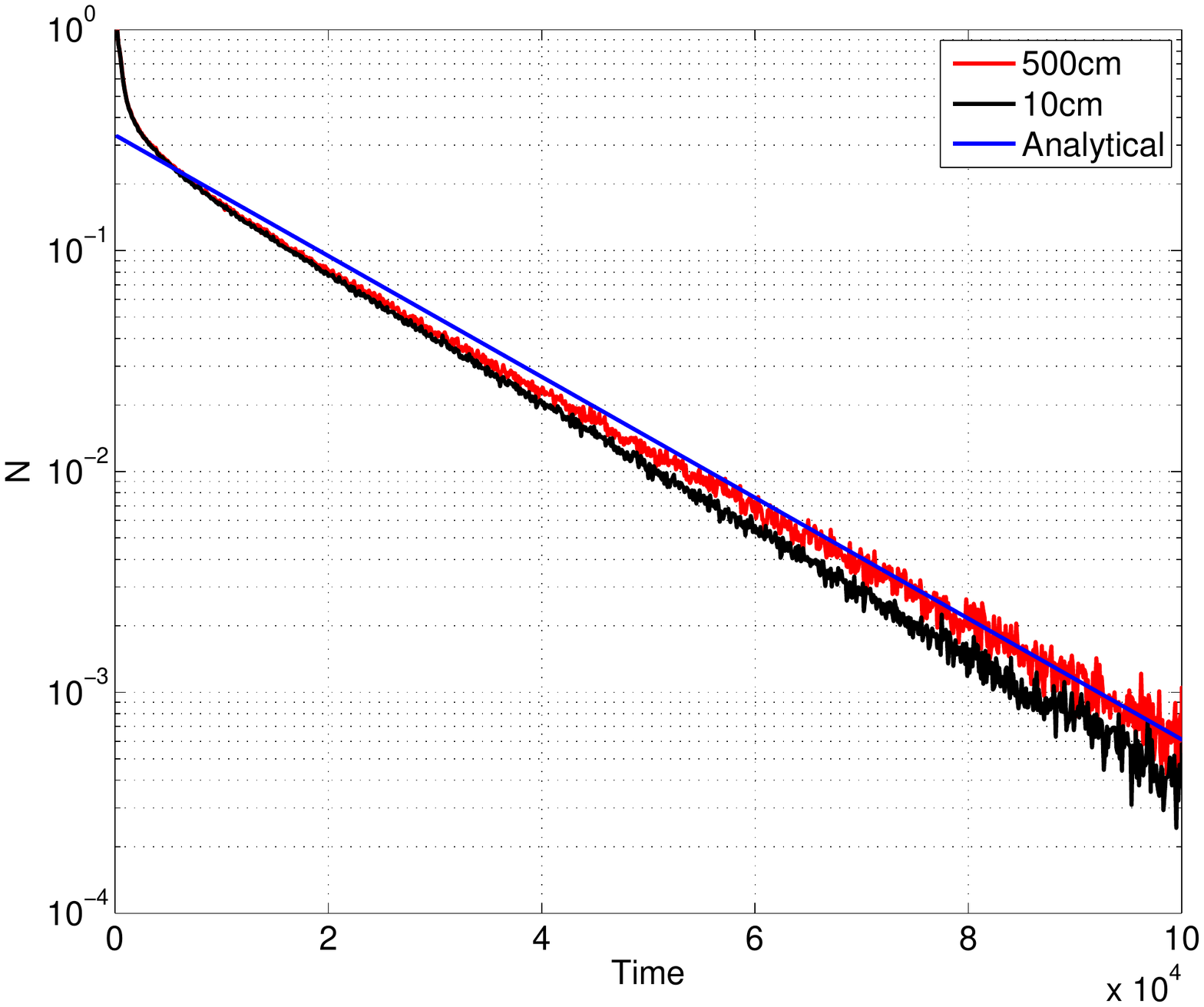}
\end{minipage}
\hspace{0.5cm}
\begin{minipage}[b]{0.5\linewidth}
\centering
\includegraphics[bb=0.0in 3.0in 10in 8in, width=7cm, height = 4cm]{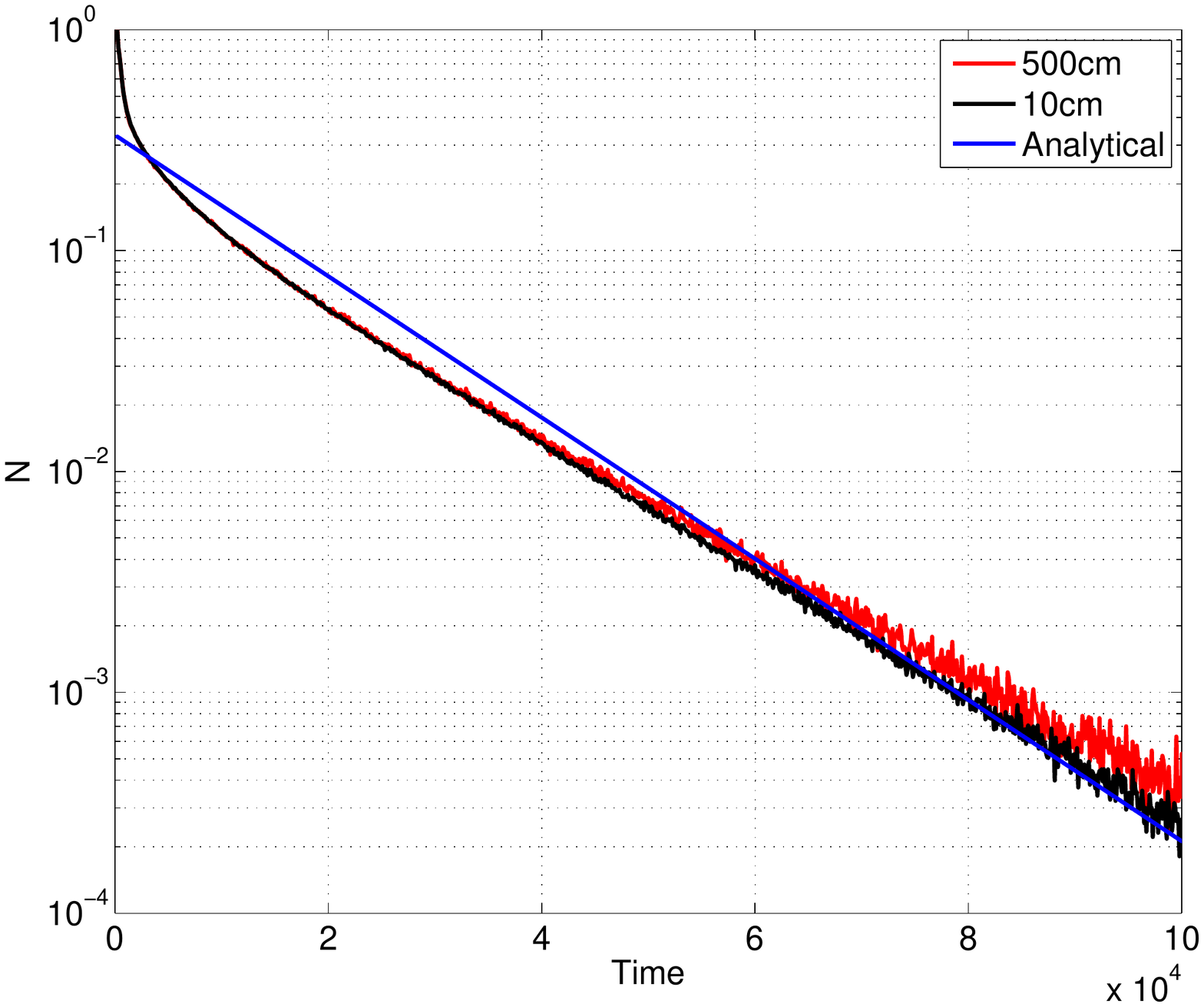}
\end{minipage}
\caption{Fits for analytical expression for $\omega_1 = 0.6318$ to the DDAA (left) and $\omega_1 = 0.9578$ to the DDSI (right).}
\label{fig:fig2}
\end{figure}

\begin{figure}[ht!]
\begin{minipage}[b]{0.5\linewidth}
\centering
\includegraphics[bb=0.5in 3.0in 9.5in 8in, width=7cm, height = 3.65cm]{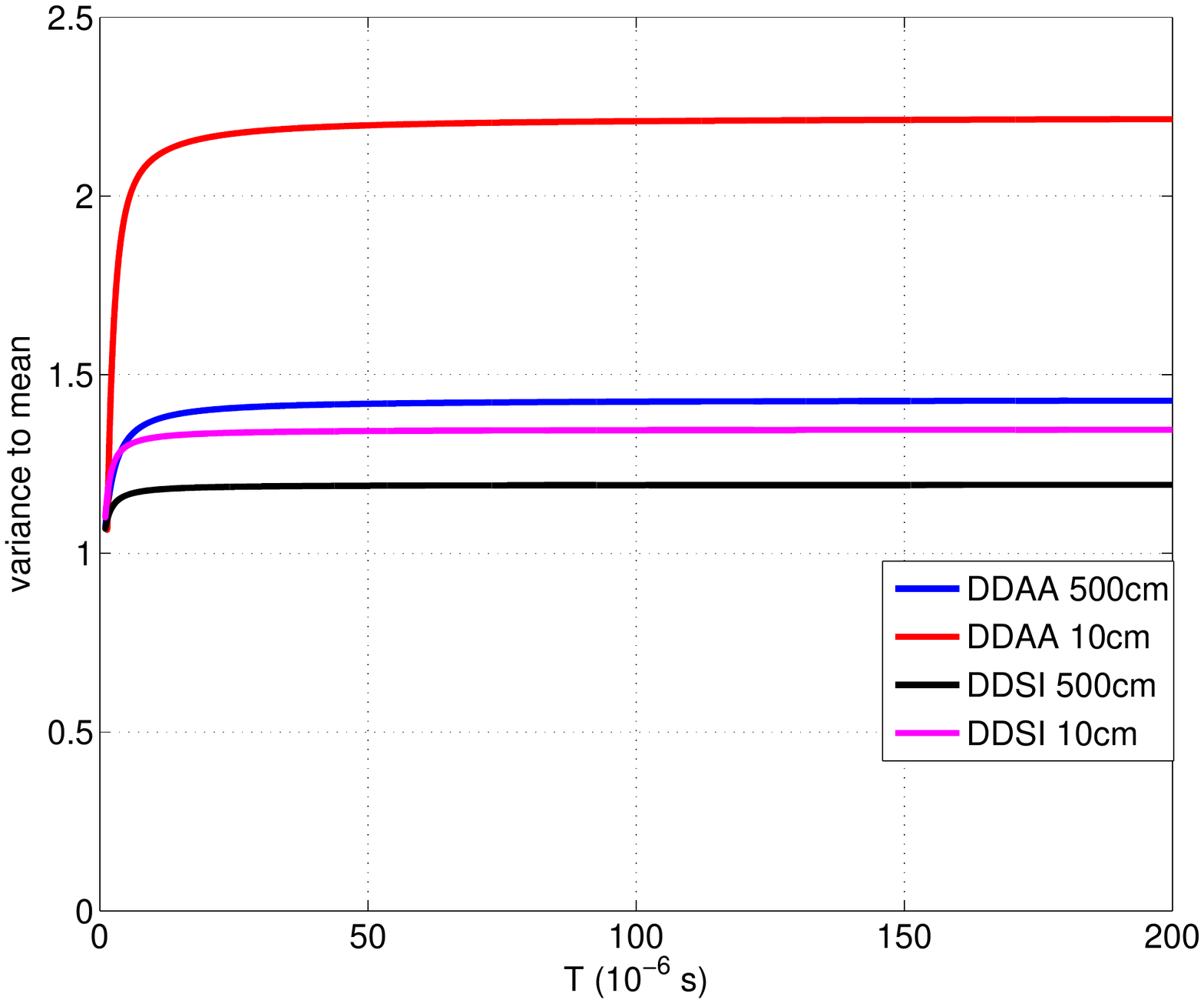}
\end{minipage}
\begin{minipage}[b]{0.5\linewidth}
\centering
\includegraphics[bb=1.0in 4.0in 9in 8in, width=7cm, height = 4cm]{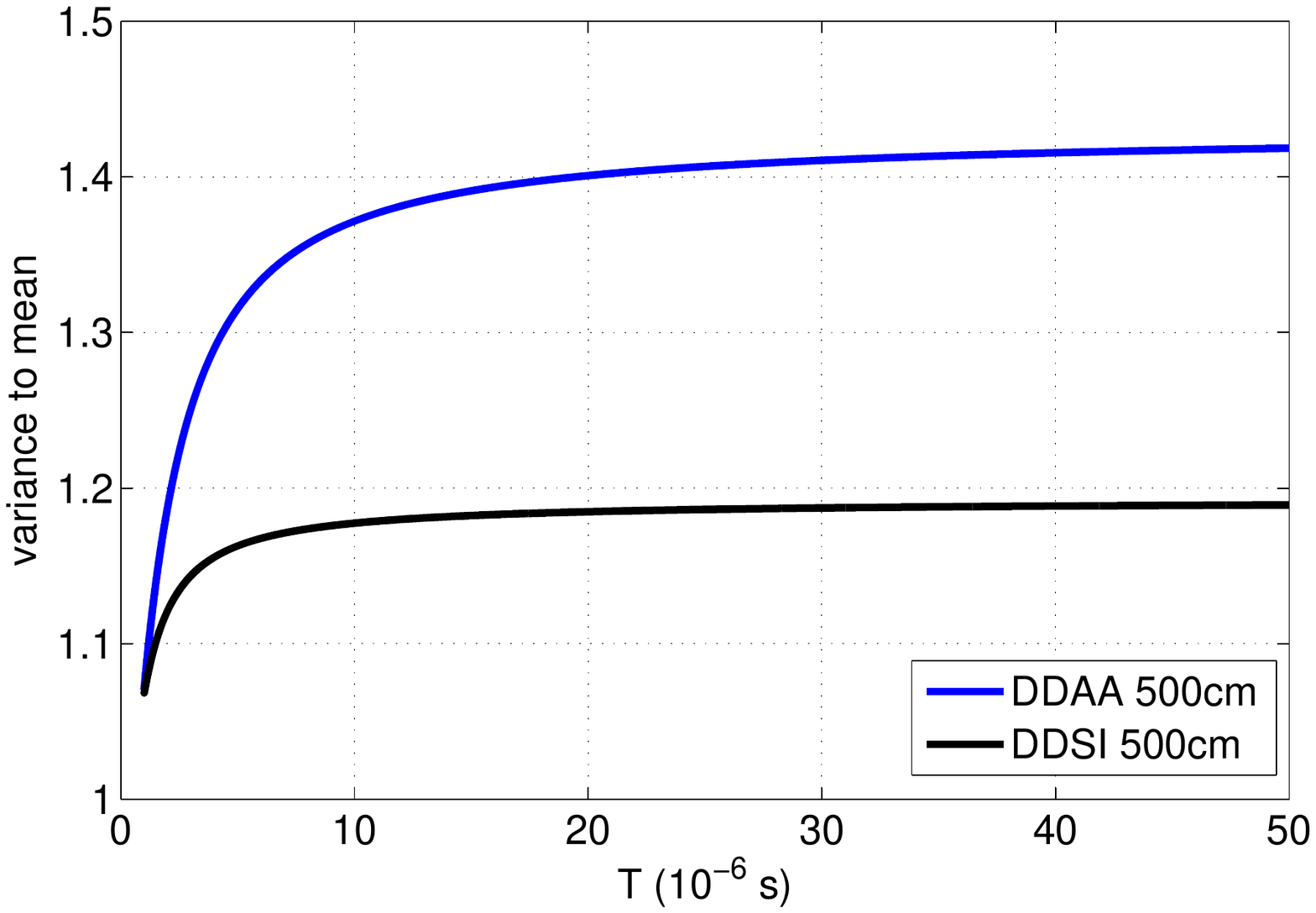}
\end{minipage}
\caption{Evaluation of the variance to mean using the values of the reaction intensities in Table. \ref{table:simulat} for (left figure) DDAA, 500cm (blue line) and 10cm (red line) and DDSI 500cm (black line) and 10cm (magenta line); and for (right figure) DDAA, 500cm (blue line) and DDSI 500cm (black line).}
\label{fig:fig3}

\end{figure}
In order to evaluate analytical and numerical results for DDSI and DDAA cases, time distributions of the current of neutrons with energy higher than 0.3 MeV crossing surface 1 (region I) in the direction of region II (tally F1 - MCNPX) were calculated. In Figure 2, one can see a comparison of the analytically estimated $\omega_1$ using Eq. (\ref{eq:1.13}) in the infinitely thick moderator case and the simulation data of the neutron rates for DDAA (left) with $\omega_1 = 0.6318$ and DDSI (right) with $\omega_1 = 0.9578$. Good agreement is found between the numerical simulation and the analytical result.

The variance to mean expression in Eq. (\ref{eq:1.27}) is displayed in Figure 3 using the data from Table 1 whereas the remaining parameters are $\nu_1 = 2.80$, $\nu_2 = 4.635$, $\lambda_d = 0.1$, $r_1 = 1.0$ and $r_2 = 0.0$. The results are shown for DDAA, 500cm (blue line) and 10cm (red line) and DDSI 500cm (black line) and 10cm (magenta line). The reaction intensities of the moderated cases are approximately of the same order of magnitude and thus the two exponentials are not easily distinguished. We find that there is a difference between the variance to mean for the DDAA and DDSI cases where the DDAA gives elevated values. For both DDAA and DDSI the infinitely moderated cases have lower values compared to the cases with finite moderation.

\section{Discussion and conclusions}
We have developed a forward Kolmogorov approach for the two point theory of the Differential Die-Away Self Interrogation variance to mean (DSVM), including a compound Poisson source and the detection process. The results are an extension to those reported in \cite{a1, a2, anderson2012} where only one finite domain was considered. However, we have restricted this study to be energy-independent and thus obtaining simplified analytical results. Retaining a two-point two-region theory, we expect to find up to four distinct exponentials. Furthermore, we have used Monte Carlo (MCNPX) simulations to find the reaction intensities needed to quantitatively assess the DSVM formula. We find that there is a difference between the variance to mean for the DDAA and DDSI cases where the DDAA gives elevated values in the limit of long gate times. For both DDAA and DDSI the infinitely moderated cases have lower values compared to the cases with finite moderation. In agreement with the results in \cite{anderson2012}, we find that the presence of two exponents in the variance to mean (due to reaction intensities that are of the same order of magnitude) is most often not clearly visible. On one hand, this indicates that detection of the presence of fissile material may not be as obvious as with the Rossi-alpha method. On the other hand, two indistinguishable exponents in the variance to mean is not a problem in itself since determining $\omega_1$ and $\omega_2$ by curvefitting could be more accurate. Elucidating on the diagnostic value of the exponents in terms of determination of the sample parameters is not clear yet, and it requires further investigations, which will be reported in future work.
\section{Acknowledgements}
This work was supported by the Swedish Radiation Safety Authority (SSM) and the 7th EU FP project FREYA.
\appendix
\thispagestyle{empty}
\setcounter{equation}{0}
\section{The Feynman-alpha formula}
In this appendix we give the full expressions of the $Y_1$ and $Y_2$ in Eq. (\ref{eq:1.27}),
\begin{eqnarray}
Y_1 & = & \nu_2 \frac{\lambda_d (\lambda_2 - \nu_1^2 \lambda_{2f}) \lambda_{T1} \lambda_{T2}}
{\omega_1^2 \omega_2^2} \frac{\omega_1}{\omega_1 + \omega_2} y_1, \\
Y_2 & = & \nu_2 \frac{\lambda_d (\lambda_2 - \nu_1^2 \lambda_{2f}) \lambda_{T1} \lambda_{T2}}
{\omega_1^2 \omega_2^2} \frac{\omega_2}{\omega_1 + \omega_2} y_2.
\end{eqnarray}
Where $y_1$ and $y_2$ are found by,
\begin{eqnarray}
y_1 & = & \frac{1}{2} \frac{2 (\lambda_2 - \nu_1^2 \lambda_{2f})^3 (\omega_2 - \lambda_1)(\omega_1 + \omega_2) - (\lambda_{T1} \lambda_{T2})^2 \omega_2}{ (\lambda_2 - \nu_1^2 \lambda_{2f})^2 (\omega_2 - \omega_1) \omega_1 \omega_2} \nonumber \\
& + & \frac{\lambda_{T1} \lambda_{T2} (\lambda_2 - \nu_1^2 \lambda_{2f})( (\lambda_1 - \nu_1^1 \lambda_{1f})(\omega_2 - \omega_1)}{(\lambda_2 - \nu_1^2 \lambda_{2f})^2 (\omega_2 - \omega_1) \omega_1 \omega_2} \nonumber \\
& + & \frac{(\omega_1+\omega_2)((\lambda_2 - \nu_1^2 \lambda_{2f}) + \omega_1 + 3 \omega_2))}{(\lambda_2 - \nu_1^2 \lambda_{2f})^2 (\omega_2 - \omega_1) \omega_1 \omega_2},
\end{eqnarray}
and
\begin{eqnarray}
y_2 & = & \frac{1}{2} \frac{2 (\lambda_2 - \nu_1^2 \lambda_{2f})^3 (\omega_2 - (\lambda_2 - \nu_1^2 \lambda_{2f}))(\omega_1 + \omega_2) + (\lambda_{T1} \lambda_{T2})^2 \omega_1}{ (\lambda_2 - \nu_1^2 \lambda_{2f})^2 (\omega_2 - \omega_1) \omega_1 \omega_2} \nonumber \\
& + & \frac{\lambda_{T1} \lambda_{T2} (\lambda_2 - \nu_1^2 \lambda_{2f}) ((\lambda_1 - \nu_1^1 \lambda_{1f}) (\omega_2 - \omega_1)}{(\lambda_2 - \nu_1^2 \lambda_{2f})^2 (\omega_2 - \omega_1) \omega_1 \omega_2} \nonumber \\
& - & \frac{(\omega_1+\omega_2)((\lambda_2 - \nu_1^2 \lambda_{2f}) + 3 \omega_1 + \omega_2))}{(\lambda_2 - \nu_1^2 \lambda_{2f})^2 (\omega_2 - \omega_1) \omega_1 \omega_2}.
\end{eqnarray}

\end{document}